\documentclass[aip,jcp,reprint]{revtex4-1}
\pdfoutput=1

\usepackage[margin=0.6in]{geometry}
\usepackage{graphicx}
\usepackage{multirow}
\usepackage{color}
\usepackage{xspace}
\usepackage{amsmath}
\usepackage[table]{xcolor}

\newcommand{\sci}[2]{\ensuremath{#1 \times 10^{#2}}}
\newcommand{\sub}[1]{\ensuremath{_{\mathrm{#1}}}}
\newcommand{\super}[1]{\ensuremath{^{\mathrm{#1}}}}

\begin{document}
\title{Improving Accuracy of Electrochemical Capacitance and\\Solvation Energetics in First-Principles Calculations}

\author{Ravishankar Sundararaman}\email{sundar@rpi.edu}
\affiliation{Department of Materials Science and Engineering, Rensselaer Polytechnic Institute, 110 8th St, Troy, NY 12180 (USA)}

\author{Kendra Letchworth-Weaver}
\affiliation{Center for Nanoscale Materials, Argonne National Laboratory, 9700 S. Cass Ave, Argonne, IL, 60439 (USA)}

\author{Kathleen Schwarz}\email{kas4@nist.gov}
\affiliation{National Institute of Standards and Technology, Material Measurement Laboratory, 100 Bureau Dr, Gaithersburg, MD, 20899 (USA)}

\begin{abstract}
Reliable first-principles calculations of electrochemical processes require accurate
prediction of the interfacial capacitance, a challenge for current computationally-efficient continuum solvation methodologies.
We develop a model for the double layer of a metallic electrode that reproduces the features of the experimental 
capacitance of Ag(100) in a non-adsorbing, aqueous electrolyte, including a broad hump in the capacitance near the Potential of Zero Charge (PZC),
and a dip in the capacitance under conditions of low ionic strength.
Using this model, we identify the necessary characteristics of a solvation model suitable for first-principles electrochemistry
of metal surfaces in non-adsorbing, aqueous electrolytes:
dielectric and ionic nonlinearity, and a dielectric-only region at the interface.
The dielectric nonlinearity, caused by the saturation of dipole rotational response in water,
creates the capacitance hump, while ionic nonlinearity, caused by the compactness of the diffuse layer,
generates the capacitance dip seen at low ionic strength.
We show that none of the previously developed solvation models simultaneously meet all these criteria.
We design the Nonlinear Electrochemical Soft-Sphere solvation model (NESS) which both captures the
capacitance features observed experimentally, and serves as a general-purpose continuum solvation model.
\end{abstract}

\maketitle

\section{Introduction}

The electrochemical double layer plays a central role in the chemistry of a wide variety
of processes including electrocatalysis, biocatalysis, and corrosion.
Over the last century, the fundamental structure and properties of the
electrochemical double layer have been characterized extensively, facilitated
by advances in spectroscopy~\cite{Boxer,Raman}, microscopy~\cite{Behm} and single-crystal electrochemistry\cite{Clavilier}.
The charge and electric field distributions in the double layer vary with applied potential in complex ways,
exhibiting a rich interplay between the surface structure, electrolyte and adsorbates.\cite{Prendergast}
Measuring, understanding, and quantitatively modeling these distributions is vital
because they can strongly affect chemical reactions at the electrochemical interface.

The charge and electric field distributions collectively contribute to the overall change in charge
with potential, or electrochemical capacitance of the double layer.
This capacitance frequently contains signatures of ion adsorption and chemical reactions
at the interface. Fundamental studies of the intrinsic response of the double layer
therefore employ single-crystalline metal surfaces and a non-adsorbing electrolyte.
Fig.~\ref{fig:capOld} shows the experimental capacitance
curves of Ag(100) in aqueous KPF\sub{6} electrolyte,\cite{Valette}
the archetypal capacitance of a metal surface in non-adsorbing aqueous electrolyte.
These curves exhibit a characteristic wide `hump' shape that is nearly independent
of electrolyte concentration, and a narrower `dip' at the potential of zero charge (PZC)
that becomes increasingly pronounced at low electrolyte concentrations.\cite{Hamelin,Parsons,TrasattiReview,Frumkin}

The concentration-dependent dip in the electrochemical capacitance can be
explained by the Gouy-Chapman-Stern (GCS) model\cite{Stern}, one of the simplest
models for the interface. This model contains two capacitance contributions in series:
a concentration-independent contribution from the dielectric region devoid of ions
next to the electrode (Helmholtz capacitance), and the strongly concentration
and potential-dependent contribution from ions that are further away (diffuse capacitance).
The latter ionic contribution is strongly nonlinear with the potential and
grows exponentially as the potential deviates from the neutral value,
producing the dip feature in the net series capacitance.
However, this model does not exhibit the reduction in capacitance
further away from the potential of zero charge (the hump).

More generally, the total series capacitance observed experimentally can be decomposed into 
a diffuse contribution, and a concentration-independent but potential-dependent
contribution containing the hump.\cite{Grahame,ParsonsZobel}
Theories to explain this hump have historically included dielectric saturation of water,
 adsorption of ions, and ion packing.\cite{TrasattiReview,Frumkin,Schmickler,ElectrifiedInterface,Badiali, MPBtheory}

Capturing complex capacitance behavior in calculations
that include electronic structure effects of the surface and the reaction species
is essential for understanding double layer effects on adsorbates and chemical reactions.\cite{Goddard,HeadGordon,HeadGordon2}
However, accurately capturing capacitance features with strictly {\it ab initio} methods
would require {\it ab initio} molecular dynamics (AIMD) calculations that are unfeasibly large
because of the large number of solvent molecules and electrolyte ions necessary for
meaningful statistics, and the long equilibration time of the electrolyte distribution.
A tractable alternative is to couple density-functional theory (DFT) calculations of the electronic-structure
sensitive reactants and surface with a continuum representation of the electrolyte.
A continuum solvation model suitable for this purpose must then
capture the effect of non-adsorbing electrolytes, while the electrode
and specific adsorbates are included explicitly in the DFT calculation.

In this paper, we demonstrate the shortcomings of current state-of-the-art
solvation models for metallic electrodes in non-adsorbing, aqueous electrolyte,
and find that they do not reproduce
the capacitance features that are experimentally observed.
We then use a simple extension of the Gouy-Chapman-Stern model to
illustrate the need for a dielectric-only region at the interface,
and nonlinear dielectric effects in the bulk electrolyte.
We formulate the Nonlinear Electrochemical Soft-Sphere (NESS)
solvation model, an extension of the recently-developed 
Soft-Sphere (atomic-sphere-based) solvation model\cite{PCM-SoftSphere},
that includes the necessary nonlinear effects and dielectric-only region.
We show that NESS, used in conjunction with DFT calculations of the metal surface,
can capture the magnitude, shape, and ionic concentration dependence of the measured
electrochemical capacitance of Ag(100) immersed in aqueous KPF\sub{6}, while
simultaneously remaining accurate for solvation energies of ions and neutral molecules.

\section{Current solvation models}

Continuum solvation models approximate the effect of a liquid environment by placing
the solute (treated quantum-mechanically using DFT) in a cavity within a dielectric continuum.
Ions in an electrolyte additionally contribute a Debye screening response in the continuum.
Beyond this ansatz, solvation models differ in the construction and parametrization
of the cavity, as well as the treatment of the continuum dielectric and ionic response.
Broadly, determining the cavity from the solute electron density (iso-density)
and/or electrostatic potential involves a few easily-determined parameters,
whereas determining the cavity using atom-centered spheres involves
many parameters and more complexity but offers greater flexibility.
Additionally, the approximations used to model the dielectric and ionic continuum
regions can be either linear or nonlinear in response to the applied potential.
Nonlinearity of the dielectric response arises from saturation of solvent dipoles aligning to the local electric field,
while nonlinearity of the ionic response arises from the Boltzmann occupation factor of the ions at the local potential.
The latter exponential increase in local ion concentration results in increased localization
of ions near the electrode and increased diffuse contribution to the capacitance,
giving rise to the capacitance dip as discussed for GCS theory above.
Finally, the ionic and dielectric cavities can be separated by a distance typically
denoted by $x_2$ in GCS theory, or they can share the same cavity ($x_2$=0).

\begin{table}
\setlength\tabcolsep{3pt}
\begin{tabular}{|c|cc|cc|}
\hline
\multirow{2}{*}{Model} & \multicolumn{2}{c|}{Response} & \multicolumn{2}{c|}{Cavities} \\
& $\epsilon$ & $\kappa$ & Type & $x_2\ne 0$ \\
\hline
LinearPCM\cite{NonlinearPCM,PCM-Kendra}
=VASPsol\cite{VASPsol}                & L  & L  & $n$  & No  \\
CANDLE\cite{CANDLE}                   & L  & L  & $n,\phi$  & No  \\ 
SCCS\cite{PCM-SCCS,PCM-SCCS-charged}  & L  & L  & $n$  & No  \\
Anderson and Jinnouchi\cite{jinnouchi} & L  & NL & $n$  & Yes \\
Dabo et al. 2010\cite{Dabo}   & L  & NL & $n$ & Yes  \\
NonlinearPCM\cite{NonlinearPCM,Refit} & NL & NL & $n$  & No  \\
Soft-sphere\cite{PCM-SoftSphere}      & L  & none  & Atom & No \\
NESS [this work]                      & NL & NL & Atom & Yes \\
\hline
\end{tabular}
\caption{Categorization of solvation models for electrochemistry by
linearity of fluid dielectric ($\epsilon$) and ionic ($\kappa$) response,
cavity parametrization type, and presence of a separate ionic cavity ($x_2\ne 0$).
Here L = linear, NL = nonlinear, $n$ = solute electron density (iso-density models),
$\phi$ = solute electrostatic potential and Atom = atomic spheres.
\label{tab:cat}}
\end{table}

Table~\ref{tab:cat} summarizes the above choices for the predominant
solvation models used for electrochemistry, most of which adopt
linear dielectric and ionic responses, precluding treatment
of the capacitance dip and hump, as we show below.
Additionally, simultaneous accuracy for properties of molecules, ions and surfaces
is challenging in iso-density models, because of different ideal cavity sizes for
capturing solvation energies of positive and negative solute charges,\cite{CANDLE}
and for capturing capacitance of metallic surfaces.\cite{Refit}
The recent Soft-Sphere solvation model\cite{PCM-SoftSphere} brings
the added flexibility of atomic parametrization, traditionally
employed only in finite molecular calculations,\cite{PCM-review,PCM-SMD}
to calculations for periodic systems including solid-liquid interfaces.
This model has so far not included ionic response or nonlinearity,
and has not yet been applied to electrochemical systems.
We take advantage of the flexibility of the atomic parametrization of the
Soft-Sphere model, and add the key ingredients required for accurate
electrochemical predictions that we show below: a separate ionic cavity
($x_2\ne 0$) and nonlinear responses of the dielectric and ions,
resulting in the NESS model.

We start by comparing the electrochemical capacitance of Ag(100) in non-adsorbing
KPF\sub{6} electrolyte against the capacitance predicted by previously-developed solvation models.\cite{Valette}
For each solvation model, we perform a series of grand-canonical DFT calculations\cite{GC-DFT}
in the JDFTx plane-wave DFT software\cite{JDFTx} for 25 uniformly spaced potentials
in a range extending 0.8~V on either side of the neutral potential (PZC),
and calculate the differential capacitance as a finite-difference derivative.
We use a seven-layer Ag(100) slab with $12\times 12\times 1$ $k$-point sampling,
cold smearing with width $\sigma = 0.01~E_h$ (see SI for convergence checks),
the `PBE' exchange-correlation functional,\cite{PBE} and `GBRV' ultrasoft
pseudopotentials\cite{GBRV} with a 20~$E_h$ (Hartree) plane-wave cutoff.
We set the vacuum/electrolyte spacing to at least three times the Debye
screening length for each ionic concentration, additionally employing truncated Coulomb
interactions\cite{TruncatedEXX} to isolate periodic images along the third direction.\cite{Dabo}

\begin{figure}
\includegraphics[width=\columnwidth]{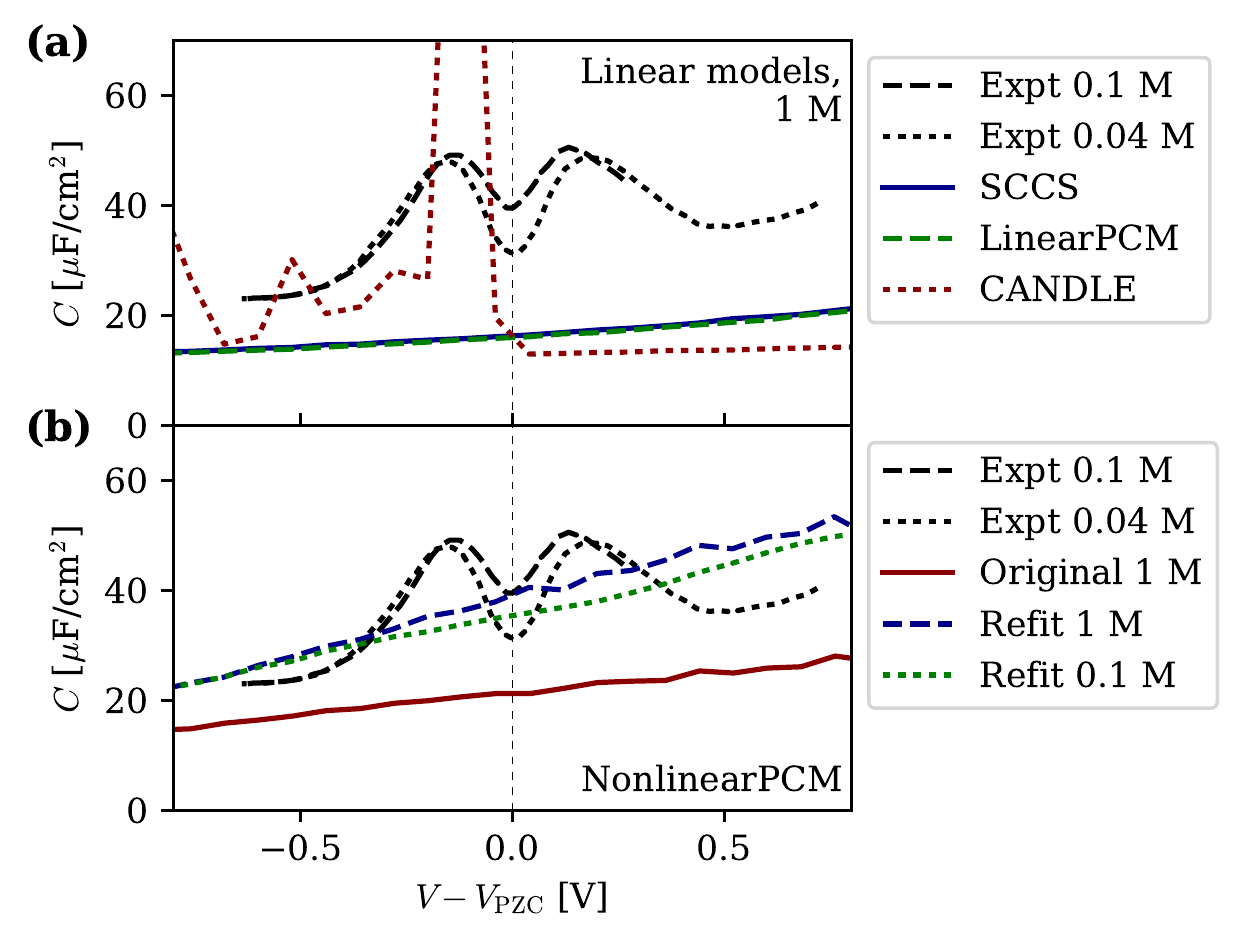}
\caption{Comparison of capacitance of Ag(100) in KPF$_6$ with
experimental measurements\cite{Valette} for
(a) linear solvation models including SCCS\cite{PCM-SCCS},
LinearPCM\cite{NonlinearPCM} and CANDLE\cite{CANDLE} at
a fixed ionic concentration of 1 M (M = mol/L), and
(b) nonlinear solvation models including the original parametrization
of NonlinearPCM\cite{NonlinearPCM} and its refit version.\cite{Refit}
The linear models show none of the potential or concentration dependence features
of the experimental data,\cite{Refit} while refit NonlinearPCM exhibits only 
a modest capacitance reduction near the PZC at low ionic concentration.
\label{fig:capOld}}
\end{figure}

First, Fig.~\ref{fig:capOld}(a) compares the experimental capacitance of Ag(100)
with solvation models that solve the linearized Poisson-Boltzmann equation.
Of these, the SCCS\cite{PCM-SCCS} and LinearPCM\cite{NonlinearPCM} models
lead to a capacitance well below the experimental values, 
with a slight linear increase with potential,
capturing none of the potential or
electrolyte-concentration dependence, as expected and discussed previously.\cite{Refit}
(Note that the commonly used VASPsol model\cite{VASPsol} re-implements
and hence produces identical results to LinearPCM.\cite{NonlinearPCM})
The CANDLE~\cite{CANDLE} model exhibits higher capacitance for negative potentials,
along with an unphysical spike when transitioning from positive to negative potentials,
because it adjusts the cavity size based on solute electric field to account for
asymmetry in solvation of cations and anions. (See SI for a more detailed discussion on the
capacitance of CANDLE.)

Next, Fig.~\ref{fig:capOld}(b) shows the effect of including dielectric saturation
from the rotational response of water molecules as well as the response
of electrolyte ions based on the nonlinear Poisson-Boltzmann equation, as in the NonlinearPCM model.\cite{NonlinearPCM}
The overall capacitance of this model is too low because of a large cavity
that results in too low a Helmholtz capacitance, which also dominates the
series capacitance and masks nonlinear effects in the diffuse capacitance.
Refitting this model's cavity to match the typical magnitude of the
experimental capacitance\cite{Refit} exhibits
some reduction in capacitance
near the PZC for lower ionic concentrations.
However, the capacitance still varies almost linearly with potential, with a larger positive slope.
This asymmetry with respect to the potential of zero charge results from the electron-density
parametrization of the solvent cavity: at positive potentials, the positively charged
surface has a lower electron density tail extending from the surface,
which pulls the cavity closer and increases the capacitance. \cite{PCM-Kendra}
This is a particularly large effect for the refit version of NonlinearPCM,
because the cavity is determined by a higher electron density threshold,
which occurs relatively closer to the metal atoms.

\section{Toy model}

\begin{figure}
\includegraphics[width=\columnwidth]{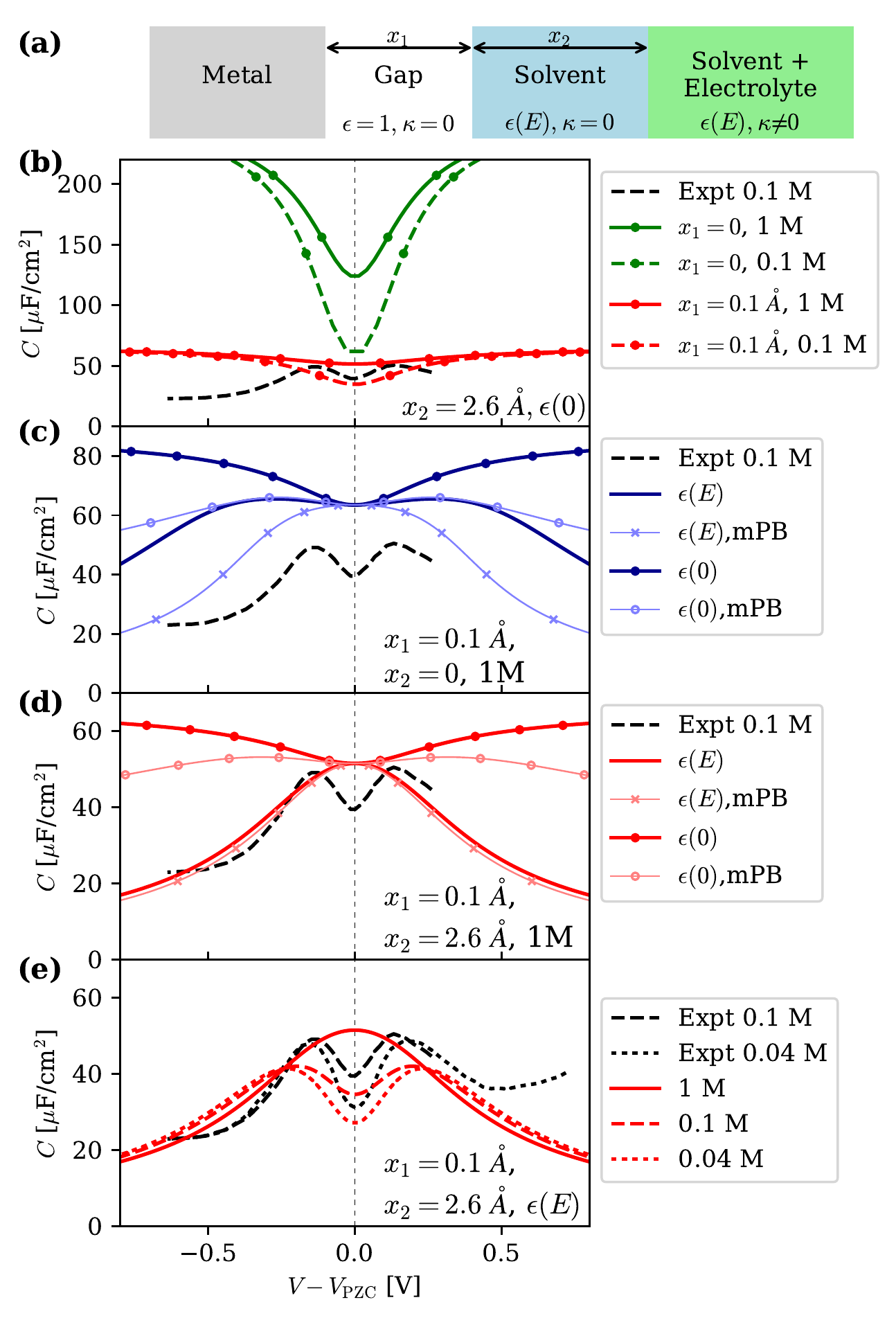}
\caption{Experimental capacitance of Ag(100)\cite{Valette}
compared against the toy model illustrated in (a) containing
vacuum and solvent gaps of width $x_1$ and $x_2$ respectively
between the metal electrode and electrolyte.
(b) Effect of $x_1$ on model capacitance for a linear
dielectric $\epsilon(0)$, where $x_1=0$ is GCS theory.
(c,d) Effect of nonlinearity ($\epsilon(E)$) and ion packing as described by the modified Poisson-Boltzmann
(mPB) theory on model capacitance for (c) $x_2=0$ and (d) $x_2=2.6$~\AA~;
both nonzero solvent gap and dielectric saturation are needed to match experimental shape
and mPB theory does not produce a hump to the extent observed in experiment.
(e) Electrolyte concentration dependence of the capacitance:
nonlinear toy model with $x_1$ and $x_2$ reproduces hump and dip
features well.
\label{fig:capToy}}
\end{figure}

To understand the limitations of existing DFT-based solvation approaches and identify
the key pieces necessary to correctly capture electrochemical capacitance,
we next study a model system illustrated in Fig.~\ref{fig:capToy}(a) that minimally extends GCS theory.
The model is composed of a continuum plane of charge representing the metal, a continuum solvent,
and a continuum electrolyte.
The metal and solvent are separated by a small gap of width $x_1$,
and the ions are excluded from a solvent region of width $x_2$.
Assuming the metal is at potential $\phi_0$ and extends until $x=0$,
the potential $\phi(x)$ for $x>0$ satisfies the nonlinear Poisson-Boltzmann equation
\begin{flalign}
\nabla\cdot\epsilon(x,|\nabla\phi|)\nabla\phi &+ \kappa^2(x,\phi) \phi = 0,\label{eq:Toy}\\
\textrm{with }\epsilon(x,E) \equiv&\ 1+\Theta(x-x_1)(\epsilon(E)-1),\\
\kappa^2(x,\phi) \equiv&\ \Theta(x-x_1-x_2)\frac{8\pi N\sub{ion}e^2\sinh\frac{\phi}{k_BT}}{\phi}
\end{flalign}
and with boundary conditions $\phi(0)=\phi_0$ and $\phi(\infty)=0$.
Here, $\epsilon(E)$ is the field-dependent dielectric constant of liquid water,
$N\sub{ion}$ is the concentration of a 1:1 electrolyte,
$\kappa$ is the inverse Debye screening length, $T$ is temperature, $e$ is the elementary charge,
$k_B$ is the Boltzmann constant, and $\Theta(x)$ is the Heaviside step function.
We use the implicit form $\epsilon(E) = \epsilon_b / (1 + 4(D/D_0)^2 + 3(D/D_0)^4)$
with $D \equiv \epsilon(E)E$, bulk dielectric constant $\epsilon_b = 78.4$
and fit parameter $D_0 = 36~\mu$C/cm$^2$, which accurately reproduces
SPC/E molecular dynamics predictions for $\epsilon(E)$.\cite{PolarizableCDFT}
We solve differential equation (\ref{eq:Toy}) using a 1D finite-element method for a series of $\phi_0$,
calculate the metal surface charge density $\sigma = -4\pi
\left.\frac{\partial\phi}{\partial x}\right|_{x=0}$ for each,
and extract the differential capacitance $\partial\sigma/\partial\phi_0$
from a finite difference derivative.
In order to consider if ion packing is responsible for the hump,
we also solve the modified Poisson-Boltzmann (mPB) equation,\cite{MPBtheory}
where the effective $\kappa^2$ is instead given by
\begin{equation}
\kappa^2(x,\phi) \equiv \Theta(x-x_1-x_2)
	\frac{8\pi N\sub{ion}e^2\sinh\frac{\phi}{k_BT}}
	{\phi\left(1+4\eta\sinh^2\frac{\phi}{2k_BT}\right)}.
\end{equation}
For the ion packing fraction $\eta \equiv \frac{4\pi}{3}(R\sub{cat}^3 + R\sub{an}^3)N\sub{ion}$,
we use an upper-bound cation radius of $R\sub{cat}=2.79$~\AA~ for K$^+$
including its hydration shell\cite{IonicRadiusK}, and an anion radius
of $R\sub{an}=2.41$~\AA~ for the PF$^-_6$ ion.\cite{IonicRadiusPF6}

Fig.~\ref{fig:capToy}(b) shows the capacitance predicted by
this toy model for the case of the linear dielectric, $\epsilon(0)$.
With $x_1=0$, the model reduces to GCS theory and the predicted capacitance
is much higher than experiment for a typical value of $x_2=2.6$~\AA.
A small vacuum region, $x_1=0.1$~\AA~, is necessary to make the magnitude
of the capacitance reasonable; note that this small vacuum gap has
the same capacitance as a solvent gap $x_2=7.8$~\AA~ due to the
bulk dielectric constant of water, $\epsilon_b = 78.4$.
With a linear dielectric, GCS theory and this new model both produce a constant
capacitance far from PZC.  Fig.~\ref{fig:capToy}(c,d) demonstrate that 
introducing $\epsilon(E)$, the nonlinear dielectric
which includes dielectric saturation, produces the hump feature
(the reduction of capacitance far from PZC).
Also note that the shape of the hump agrees best with experimental measurements for a theory which includes a solvent-only region,  
characterized by non-zero $x_2=2.6$~\AA~ in Fig.~\ref{fig:capToy}(d). When $x_2=0$, as in Fig.~\ref{fig:capToy}(c), the width and the height of the capacitance hump are both too large compared to the experiment.
Fig.~\ref{fig:capToy}(c,d) also show the effects of ion packing, as described by mPB theory with a
modified $\kappa^2$ as discussed above.  Inclusion of ion packing produces a hump-like feature,
but with a substantially smaller magnitude than observed in experiment.  While
this effect may be dominant for ionic liquids~\cite{MPBtheory},
dielectric saturation is the more important effect for non-adsorbing aqueous electrolyte
because of water's high dielectric constant, the small size of the ions, and the
fact that water molecules (rather than ions) are closest to the surface.
Additionally, the NonlinearPCM solvation model \cite{NonlinearPCM} already includes 
ion packing effects at a level comparable to mPB theory \cite{MPBtheory}  
and yet it did not produce an observable hump in Fig.~\ref{fig:capOld}.

Fig.~\ref{fig:capToy}(e) shows that the nonlinear toy model
with non-zero $x_1$ and $x_2$ has both the expected capacitance maximum
and dips depending on the ionic concentration.
However, relatively minor disagreements remain with the experimental curve,
including the missing plateaus and slight asymmetry far from PZC, speculated to arise from
specific adsorption of water or ions~\cite{Trasatti,valetteKion,Frumkin}.
Understanding the origin of these secondary features seen experimentally
and capturing them in a solvation model will be the subject of future work.
For instance, future studies could
consider the interplay between dielectric saturation and electrolyte packing 
using a method such as classical density-functional theory with molecular orientations for the solvent \cite{RigidCDFT,PolarizableCDFT}.

\section{New solvation model}

Here, we focus on obtaining a continuum solvation model for DFT
that captures the hump and dip features qualitatively, and
Fig.~\ref{fig:capToy} indicates that such a solvation model must be
nonlinear in both the dielectric and ionic responses, and it must have a
solvent region that excludes ions ($x_2\ne 0$).
(Continuum solvation models already include $x_1\ne 0$ due to the gap
between the electron density of the solute and the solvent cavity.)
Table~\ref{tab:cat} illustrates that none of the previously developed solvation models
simultaneously include both nonlinearities and $x_2\ne 0$.
Additionally, our previous work~\cite{CANDLE,Refit} found that electron density-based
parametrization of solvation model cavities results in limited simultaneous accuracy
for properties of molecules, ions, and metal surfaces.

We therefore construct a solvation model with cavities based on atom-centered spheres
that is also suitable for calculations of extended systems in the plane-wave basis,
starting from the recently proposed Soft-Sphere solvation model.\cite{PCM-SoftSphere}
In this model, the cavity shape function is determined as
\begin{equation}
s(\vec{r}) = \prod_i \frac{1}{2}\mathrm{erfc}\frac{R_i-|\vec{r}-\vec{r}_i|}{\sigma}
\label{eq:sSS}
\end{equation}
which goes smoothly from zero in the vicinity of atoms located at positions $\vec{r}_i$
to one in the solvent region more than the radius $R_i$ away,
with a transition width $\sigma$ set to 0.5 Bohr.
The sphere radii $R_i = f R_i^0$, where $f$ is a scale parameter fit
to solvation energies, and $R_i^0$ are a set of standard atomic radii,
chosen to be the universal force field (UFF)~\cite{UFF} van der Waals radii in Ref.~\citenum{PCM-SoftSphere}.

The Soft-Sphere solvation model includes only linear dielectric response
and neglects Debye screening from the electrolyte and hence cannot correctly capture electrochemical capacitance.
Therefore, we construct the Nonlinear Electrochemical Soft-Sphere (NESS) solvation model
by using the Soft-Sphere cavity definition in NonlinearPCM~\cite{NonlinearPCM},
and including an additional ionic cavity so that $x_2$ is non-zero.
Specifically we set the spatial modulation $s(\vec{r})$ of the nonlinear
dielectric response in NonlinearPCM, exactly as given by (\ref{eq:sSS}),
(instead of a threshold $n_c$ on the electron density $n(\vec{r})$ as in Ref.~\citenum{NonlinearPCM}).
Further, we set the spatial modulation of the nonlinear ionic response
to a separate cavity function $s\sub{ion}(\vec{r})$, which is still given
by (\ref{eq:sSS}) except with modified radii $R_i^{\mathrm{ion}} = f R_i^0 + x_2$,
which introduces an electrolyte-excluded solvent region of width $x_2$.
All remaining details of the model are identical to NonlinearPCM
and share most of its implementation in JDFTx,\cite{JDFTx}
including the nonlinear response functions, the term with effective
surface tension $t$ to capture non-electrostatic free energy contributions
and the algorithms to converge the solvation free energy and to handle
charged solutes in periodic boundary conditions.
(See Refs.~\citenum{NonlinearPCM}~and~\citenum{GC-DFT} for further details.)

\begin{table}
\setlength\tabcolsep{2pt}
\begin{tabular}{|c|ccc|c|}
\hline
\multirow{2}{*}{Model} & \multicolumn{4}{c|}{MAE [kcal/mol]} \\\cline{2-5}
 & Neutral & Cations & Anions & All \\
\hline
\multicolumn{5}{|l|}{\bf Linear dielectric} \\
\hline
LinearPCM\cite{NonlinearPCM}=VASPsol\cite{VASPsol}
                                 & 1.27 & 2.10 & 15.1 & 3.59 \\
SCCS\cite{PCM-SCCS}              & 1.20 & 2.55 & 17.4 & 3.97 \\
CANDLE\cite{CANDLE}              & 1.27 & 2.62 & 3.46 & \textbf{1.81} \\
Soft-sphere\cite{PCM-SoftSphere} & 1.25 & 6.02 & 10.4 & 3.41 \\
\hline
\multicolumn{5}{|l|}{\bf Nonlinear dielectric} \\
\hline
Orig. NonlinearPCM\cite{NonlinearPCM} & 1.28 & 16.1 & 27.0 & 7.55 \\
Refit NonlinearPCM\cite{Refit}        & 1.44 & 14.4 & 27.6 & 7.51 \\
NESS [this work]\cellcolor{lightgray}
 & 1.29\cellcolor{lightgray} & 3.67\cellcolor{lightgray} & 9.58\cellcolor{lightgray} & 2.96\cellcolor{lightgray} \\
\hline
\end{tabular}
\caption{Comparison of mean absolute errors (MAEs) in aqueous solvation energies
of the same set of 240 neutral molecules, 51 cations and 55 anions between
several prominent grid / plane-wave based solvation models.
CANDLE yields the lowest MAE in a single parametrization due to charge-asymmetry correction,
but that leads to unphysical features in the capacitance (Fig.~\ref{fig:capOld}).
The NESS model parametrized to neutral solvation energies is reasonably accurate
for solvation energies (highlighted below), while simultaneously predicting
electrochemical capacitance in good agreement with experiment (Fig.~\ref{fig:capSS}).
(See SI for additional comparison of ionic parametrizations of the SCCS, Soft-sphere and NESS models.)
\label{tab:maeCompare}}
\end{table}

We perform solvation free energy calculations of a large set of neutral solutes, cations
 and anions, including ionic screening (1 M electrolyte) for the ion calculations.  We fit the 
empirical parameters $f$ (cavity scale factor) and $t$ 
(effective tension = $\alpha+\gamma$ in Ref.~\citenum{PCM-SoftSphere}'s notation)
to the solvation free energies.
When fit to neutral solutes, we find optimum parameters $f = 1.00$
and $t = \sci{1.02}{-5}$, which we use for subsequent calculations here.
Table~\ref{tab:maeCompare} compares the accuracy of this parametrization for
solvation energies of neutral solutes, cations and anions with that of
several previous solvation models.
(See SI for further details on the parametrization, including fits to
cation or anion solvation energies alone.)
Our new NESS model performs comparably to the linear Soft-Sphere model,
and in its parametrization to neutral solutes, is substantially more accurate
than the original NonlinearPCM.\cite{NonlinearPCM}
It still falls short of CANDLE,\cite{CANDLE} which is able to deliver the best overall accuracy
for solvation free energies within a single parametrization because it accounts for charge asymmetry.
However, that feature of CANDLE results in unphysical capacitance predictions,
as shown in Fig.~\ref{fig:capOld} and discussed above and in the SI.

\begin{figure}
\includegraphics[width=\columnwidth]{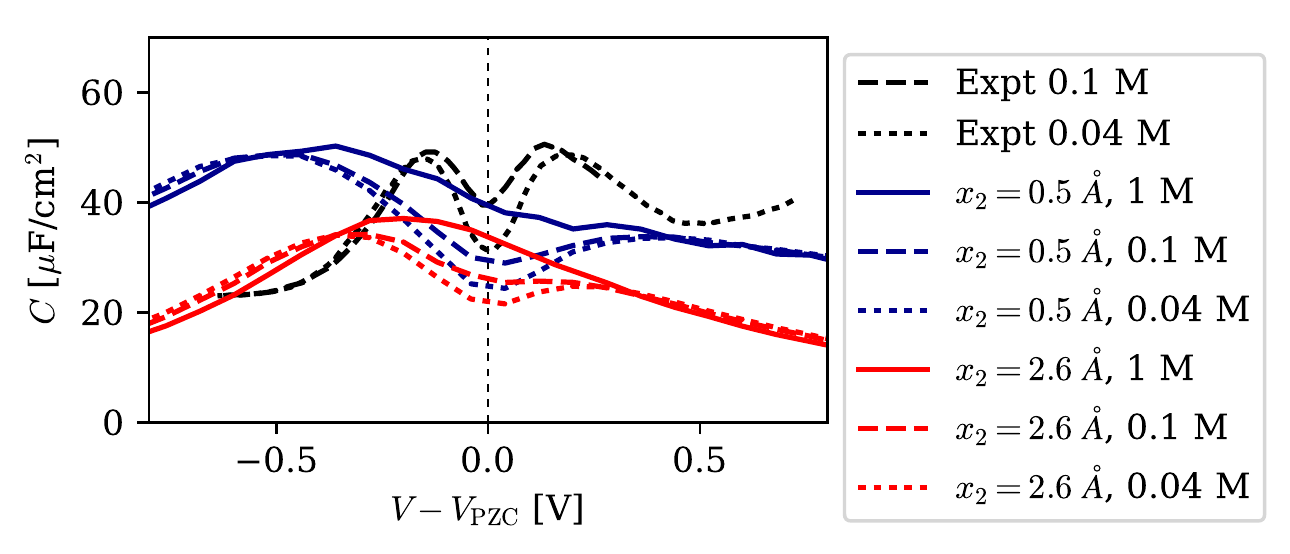}
\caption{NESS solvation model predicted capacitance for $x_2=0.5$ and 2.6~\AA,
exhibiting both the hump and dip features of the experimental capacitance.
\label{fig:capSS}}
\end{figure}

Using the new NESS model parametrization, we perform capacitance calculations
of the Ag(100) surface.  We note that for the capacitance calculations, the silver atom radius 
in the original Soft Sphere model is small enough so that the fluid leaks into the interior of the silver 
slab.  For these calculations, we have modified our model to exclude the solvent from the interior 
of the slab.  Fig.~\ref{fig:capSS} shows that the NESS model predicts
the electrochemical capacitance of Ag(100) in much better agreement
with experiment than any previous solvation model,
including both the hump and the dip features.
However, the capacitance maximum is not centered at the PZC
as was the case for the toy model in Fig.~\ref{fig:capToy}(e).
In this solvation model, the cavity is independent of the electron density,
while the electrons extend further from the metal atoms with increasing
negative charge, reducing the gap and increasing the capacitance.
As $x_2$ increases, the contribution of this gap capacitance to the total
series capacitance becomes less important, reducing the asymmetry of the hump, as seen in Fig.~\ref{fig:capSS}.
In reality, the solvent likely moves further away with increasing negative charge, due to excess electrons
(as captured by the iso-density cavities), resulting in a more symmetric hump.

\begin{figure}
\includegraphics[width=\columnwidth]{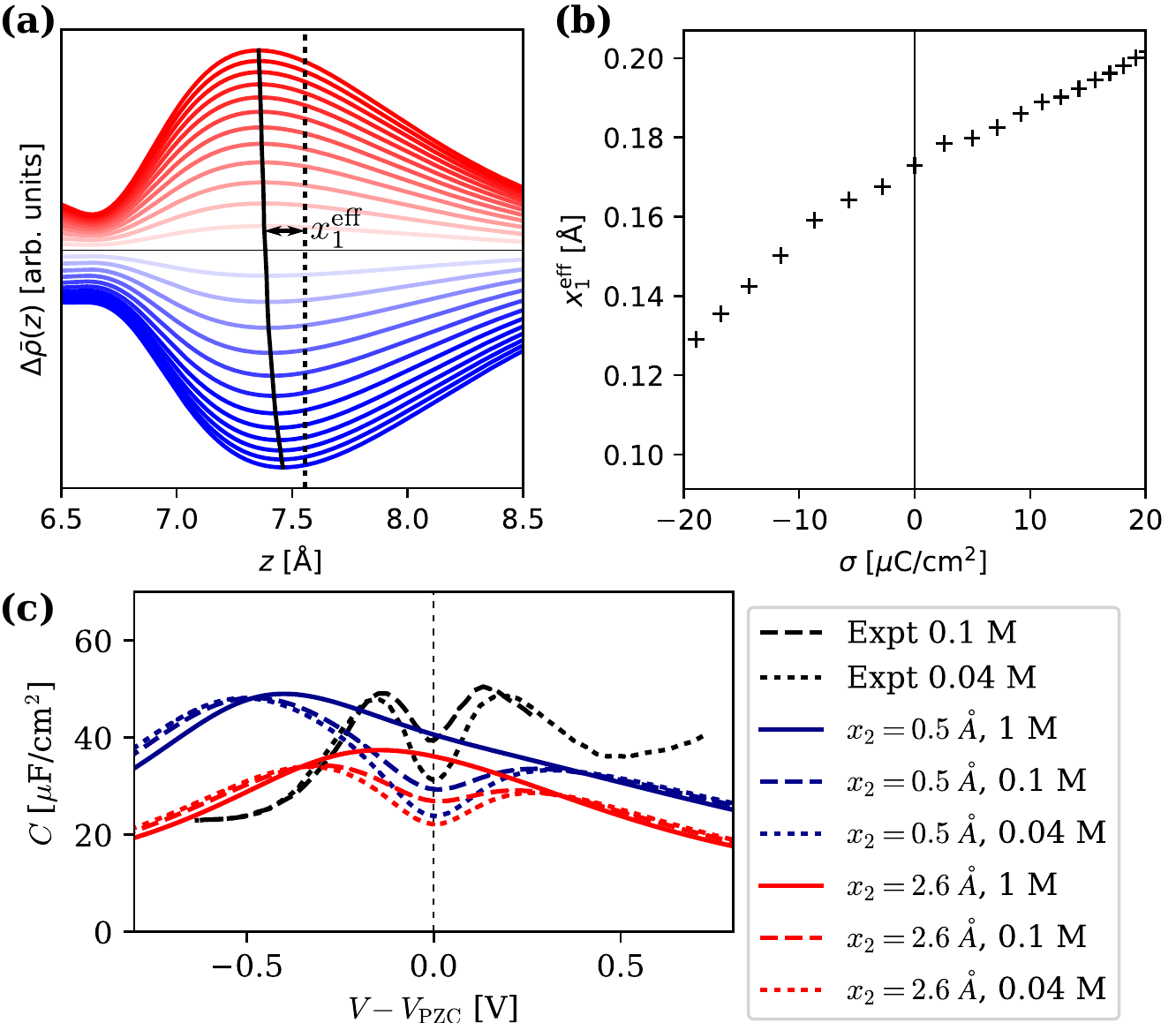}
\caption{(a) Induced charge density at Ag(100) surface as a function
of applied potential, from bottom to top: +0.8 V to -0.8 V relative to the PZC, as in Fig.~\ref{fig:capSS}.
Dotted line indicates location of dielectric cavity turn-on, and the solid line 
indicates the maximum of the induced charge density.
(b) The separation between these locations in part (a) is the effective vacuum distance,
$x_1\super{eff}$ between the metal and the solvent as defined in the toy model above,
but which now depends on the electrode surface charge density $\sigma$.
(c) Capacitance curves predicted by the toy model with this $\sigma$-dependent $x_1\super{eff}$.
\label{fig:asymmToy}}
\end{figure}

To more conclusively identify the source of the asymmetry in the capacitance curves 
from the NESS solvation model, we use the electron density results from the NESS
calculations as inputs for the $x_1$ parameter in our toy model.  Fig.~\ref{fig:asymmToy}(a)
depicts the induced charge at the Ag(100) surface as a function of applied potential,
assigning the distance between the electron density peak and the cavity onset as the $x_1\super{eff}$ parameter,
plotted in Fig.~\ref{fig:asymmToy}(b) as a function of the electrode surface charge density $\sigma$.
Fig.~\ref{fig:asymmToy}(c) depicts the results of the toy model, using the
$\sigma$-dependent $x_1\super{eff}$ results from Fig.~\ref{fig:asymmToy}(b).  
The toy model capacitance in Fig.~\ref{fig:asymmToy}(c) and the NESS capacitance of Fig.~\ref{fig:capSS}
are in very good agreement, confirming that the asymmetry in the NESS model is caused by the 
dielectric cavity remaining spatially fixed while the induced charge density shifts with potential.

\section{Conclusions and Outlook}

The NESS solvation model brings the computational electrochemistry community closer to developing an all-purpose solvation model
 for ions, molecules, and electrified surfaces, but there are still many opportunities for improvement and deeper understanding.
For instance, improvements to the dielectric cavity definition and to the dielectric response at the interface 
have potential to increase accuracy.
For the dielectric cavity, the original Soft-Sphere model and NESS employ an atomic radius for
each element, currently set based on vdW radii from the UFF force field~\cite{UFF}.
These radii have been carefully evaluated for the few atoms present in organic
solutes for which extensive solvation free energy data is readily available.
This choice fortuitously gave good results for Ag(100) capacitance,
but that might not be the case for all metal surfaces, and for all properties.
Additionally, the solvent cavity presently remains fixed with potential
and charge on the metal surface; understanding and incorporating
the variation of water-metal distance with surface charge could
result in further improvements in accuracy for the capacitance.  
We note that previous solvation models with isodensity cavity
parametrizations\cite{jinnouchi,PCM-SCCS,PCM-SCCS-charged,Gygi,CANDLE,NonlinearPCM,Refit}
mitigate some of this asymmetry, but have not yet simultaneously captured
solvation energetics and electrochemical capacitance within a single parametrization.
Therefore, future work may need to employ a hybrid approach that
combines the flexibility of atomic-sphere based cavities with the
response to electron density inherent to isodensity cavities.
Additionally, for the dielectric response at the interface and near ions,
we have assumed that the response of the water remains
unchanged from bulk; future improvements could include
modulation of the dielectric response adjacent to the solute.

In summary, we find that a solvation model must contain nonlinearities
in both the ionic and dielectric responses, along with a region of space
containing only the dielectric, in order to capture the features seen
in the experimental capacitance curves of metals such as Ag(100)
immersed in non-adsorbing electrolytes.
By including these characteristics, the NESS solvation model substantially improves
on previous solvation models for electrochemical systems, qualitatively capturing
key features in the experimental electrochemical capacitance.

\section{Supplemental Material}

See supplemental information for the convergence of the capacitance with the 
number of layers of Ag, number of $k$-points, and electron density smearing width.
Additionally, the supplemental information contains a discussion of the capacitance
behavior of the CANDLE solvation model, and parametrization details of the NESS solvation
model developed in this paper.

\section*{Acknowledgments}

RS acknowledges start-up funding from the Department of Materials Science and Engineering
at Rensselaer Polytechnic Institute and computational resources provided by the
Center for Computational Innovations (CCI) at Rensselaer Polytechnic Institute.
KLW’s use of the Center for Nanoscale Materials, an Office of Science user facility,
was supported by the U. S. Department of Energy, Office of Science, Office of 
Basic Energy Sciences, under Contract No. DE-AC02-06CH11357.

\section*{References}
\bibliographystyle{apsrev4-1}
\makeatletter{} 

\end{document}